\documentclass[fleqn,usenatbib]{mnras}  
\usepackage[T1]{fontenc}
\usepackage{ae,aecompl}

\usepackage{graphicx}	
\usepackage{amsmath}	
\usepackage{amssymb}	
\usepackage{xspace}     
\usepackage[dvipsnames]{xcolor}
\usepackage[normalem]{ulem}
\defcitealias{Hagen2019}{H19}
\defcitealias{Schonrich2019}{S19}
\defcitealias{Anders2019}{A19}

\title[The Local Velocity Ellipsoid]{The Tilt of the Local Velocity Ellipsoid as seen by Gaia}

\author[Everall et al.]{
A. Everall$^1$\thanks{Contact email: \href{mailto:asfe2@cam.ac.uk}{asfe2,nwe,vasily@cam.ac.uk}}, N.~W. Evans$^1$, V.~Belokurov$^1$, R. Sch{\"o}nrich$^2$
\\
$^{1}$Institute of Astronomy, University of Cambridge, Madingley Road, Cambridge CB3 0HA, UK\\
$^{2}$University of Oxford, Rudolf Peierls Centre for Theoretical Physics, Clarendon Laboratories, OX1 3PU Oxford, UK\\
}

\pubyear{2019}

\begin{document}
\label{firstpage}
\pagerange{\pageref{firstpage}--\pageref{lastpage}}
\maketitle

\begin{abstract}
The Gaia Radial Velocity Spectrometer (RVS) provides a sample of 7\,224\,631 stars with full six-dimensional phase space information. Bayesian distances of these stars are available from the catalogue of Sch\"onrich et al. (2019). We exploit this to map out the behaviour of the velocity ellipsoid within 5 kpc of the Sun. We find that the tilt of the disc-dominated RVS sample is accurately described by the relation $\alpha = (0.952 \pm 0.007)\arctan (|z|/R)$, where ($R,z$) are cylindrical polar coordinates. This corresponds to velocity ellipsoids close to spherical alignment (for which the normalising constant would be unity) and pointing towards the Galactic centre. Flattening of the tilt of the velocity ellipsoids is enhanced close to the plane and Galactic centre, whilst at high elevations far from the Galactic center the population is consistent with exact spherical alignment. Using the LAMOST catalogue cross-matched with Gaia DR2, we construct thin disc and halo samples of reasonable purity based on metallicity. We find that the tilt of thin disc stars straddles $\alpha = (0.909-1.038)\arctan (|z|/R)$, and of halo stars straddles $\alpha = (0.927-1.063)\arctan (|z|/R)$. We caution against the use of reciprocal parallax for distances in studies of the tilt, as this can lead to serious artefacts.
\end{abstract}


\section{Introduction}

Understanding the distribution of mass in the Milky Way is of great interest for constraining our Galaxy's formation history. Unfortunately, the majority of the mass does not emit detectable electromagnetic radiation and so we are forced to use indirect methods. One such method is to analyse the velocity dispersion of stars, as this is related to the Galactic potential through the Jeans equations. 

The sample of 7\,224\,631 stars seen by the Gaia Radial Velocity Spectrometer~\citep[hereafter RVS,][]{Brown2018,Ka19} provides a tempting dataset to study the behaviour of the velocity dispersion tensor. A recent attempt to do so was conducted by \citet[][henceforth \citetalias{Hagen2019}]{Hagen2019}. By augmenting the dataset with multiple spectroscopic surveys, including LAMOST Data Release 4 \citep[DR4,][]{LAMOST2012}, APOGEE DR14 \citep{Abolfathi2018} and RAVE DR5 \citep{Kunder2017}, \citetalias{Hagen2019} generated a sample of the Solar neighbourhood in excess of 8 million stars. They
found that the velocity ellipsoids of their sample were close to spherically aligned within the Solar radius, but became cylindrically aligned at larger radii. 

The results of \citetalias{Hagen2019} show comparable total misalignment to \cite{Binney2014} using RAVE DR5 \citep{Kunder2017}. Both studies find that the tilt of the ellipsoids of their thin disc dominated samples deviate significantly from spherical alignment in the Solar neighbourhood. The mismatch is significantly greater than found by \cite{Budenbender2015} using SEGUE G dwarfs~\citep{Yanny2009}. The disagreement is more striking when compared to the halo population. A number of studies using Sloan Digital Sky Survey data \citep{AdelmanMcCarthy2008} found an almost spherically aligned velocity ellipsoid for halo stars \citep{Smith2009,Bond2010,Evans2016}. This seems to be confirmed by the recent study of \citet{Wegg2018}, who used a set of RR Lyrae extracted from Gaia Data Release 2 to conclude that the potential of the halo is spherical. This necessarily implies that the velocity ellipsoid is spherically aligned~\citep{Smith2009,An16}. This is contrary to the results from \citetalias{Hagen2019}, where the ellipsoid is cylindrically aligned at large distances from the Galactic Center and high above the plane.

Here, we analyse the behaviour of the local velocity ellipsoid using the Gaia RVS, complemented with LAMOST. We introduce the datasets in Section 2, paying careful attention to distance errors and biases. We provide our algorithm in Section 3 and present our results in Sections 4 and 5. We find that simple use of the reciprocal of parallax as a distance estimator is dangerous and can lead to misleading results. The local velocity ellipsoid is always close to spherical alignment, and this remains true even for the thin disc and halo populations separately. The only substantial misalignment occurs for star samples at low latitudes and close to the Galactic centre, where the potential is strongly disc dominated.

\section{Data}
\label{sec:data}

\subsection{The Gaia DR2 RVS sample}

The Gaia DR2 RVS sample is a subset of the main DR2 catalogue with
radial velocities derived from the on-board spectrograph
\citep{Brown2018,Ka19}.  Although this gives us six-dimensional phase
space data for over 7 million stars, the information on the distance
is of course encoded as the parallax \citep[an introductory discussion
  how to infer distances from Gaia parallaxes can be found
  in][]{Luri2018, Bailer-Jones2015}. To recover the tilt of the
velocity ellipsoid, special care needs to be taken with the inferred
distances. Of course, to convert the proper motions into the
tangential velocities requires the distance, and so poorly computed
and noisy distances can overwhelm calculations of the tilt. We thus
face two central problems: i) the parallaxes of Gaia can be biased,
and ii) the method of inferring distances can be biased.

Concerning the parallax bias, \cite{Lindegren2018} used a sample of known quasars to determine a zero-point parallax offset of $\delta_\varpi = -29 \,\mu$as, while they also showed that the parallax uncertainties are underestimated by about $\delta \sigma_\varpi = 43 \,\mu$as, which are to be added in quadrature. The offset is known to depend on colour and apparent magnitude, might also depend on the object type and parallax, and hence is likely inappropriate for stellar objects in the RVS catalogue. More appropriate to the RVS catalogue, but still restricted to a particular subsets of stars, are a series of papers which found different parallax offsets: \citet{Riess2018} constrained $\delta_\varpi = -46 \pm 13\,\mu$as for Cepheids, whilst \citet{Xu2019} found a value of $-75 \pm 29\,\mu$as using VLBI astrometry of YSOs and pulsars. \citet{Zinn2018} and \citet{Khan2019} use asteroseismology for (mostly) red giants in the Kepler fields to get parallax offsets $\sim -50\,\mu$as, depending on the field position. For the full Gaia DR2 RVS sample, the parallax offset was calculated by \citet[][henceforth \citetalias{Schonrich2019}]{Schonrich2019} using their statistical distance method. They find an average parallax offset of $-54 \pm 0.06 \,\mu$as, where the uncertainty comprises their systematic uncertainty with a negligible statistical error.

The literature contains in principle four approaches to infer stellar distances:

\noindent
(i) Simply setting the distance $s = 1/\varpi$, as done by \citet{Hagen2019}. This approach should only be used in situations where the parallax uncertainty is negligible for the problem, since it produces a three-fold bias: neglect of the selection function, neglect of the spatial distribution of stars, and ignorance of the fact that $1/\varpi$ is not the expectation value of the probability distribution function $P(s)$. The latter bias was already identified by \cite{Stroemberg27} and became later well-known as the \citet{Lutz1973} bias. 

\noindent
(ii) Performing Bayesian distance estimates with a set of generic assumptions about the sample and the underlying Galactic density distribution, which eliminates the major problems of $s = 1/\varpi$, but leaves some uncertainties concerning the selection function. A good example of this approach is \citet{Ba18}.

\noindent
(iii) Doing a full Bayesian estimate involving stellar models, such as the \cite{Anders2019} distances.

\noindent
(iv) Doing a full Bayesian approach with a self-informed prior that estimates the selection function from the data directly \citep[][]{Schonrich2017, Schonrich2019}. 
Speaking generally, approaches (iii) and (iv) yield the most
trustworthy results, though they of course involve greater expenditure
of effort.

\begin{figure}
   \includegraphics[width=0.45\textwidth]{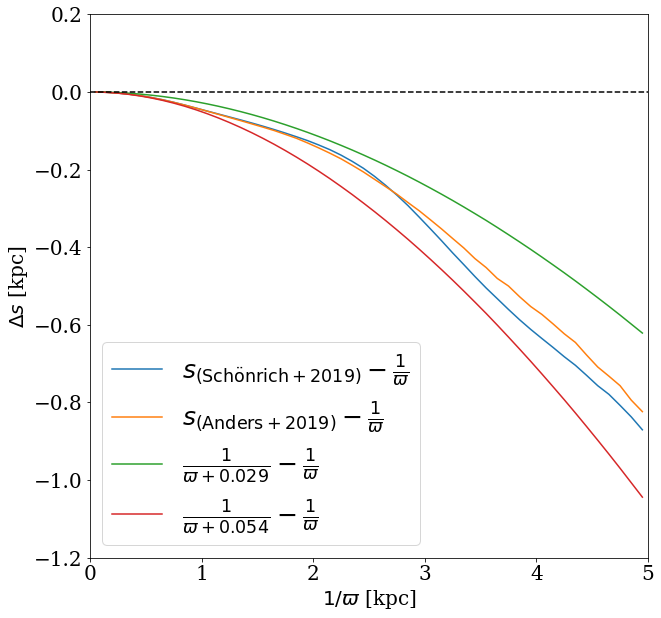}
   \caption{Running median of the distance offset from a naive parallax reciprocal. The green curve is generated by corrections using the $29\,\mu\mathrm{as}$ parallax offset suggested by \citet{Lindegren2018}, whilst the red curve uses the $54\,\mu\mathrm{as}$ parallax offset suggested by \citetalias{Schonrich2019}. Finally, the blue and orange curves show the difference between the parallax reciprocal and the Bayesian distance estimates from \citetalias{Schonrich2019} and \citetalias{Anders2019} respectively. Using the reciprocal of the parallax as a distance estimator is unwise beyond heliocentric distances $s \sim 1$ kpc.}
\label{fig:distances}
\end{figure}

The mean bias between the different $\delta_\varpi$ estimates, and distance estimators is shown in Fig.~\ref{fig:distances}. The \citetalias{Schonrich2019} distances deviate from a simple parallax reciprocal $1/\varpi$ for distances beyond $\sim 1\mathrm{kpc}$. They also show substantially greater offset than would be accounted for by the $29\,\mu$as correction. We also note that the distance deviation is smaller than if we were to use the $54\,\mu$as offset and naively use $1/\varpi$. Fig.~\ref{fig:distances} underscores the point that the crude calculation of $1/\varpi$ overestimates the distance.

Tangential velocities are calculated by multiplying the proper motion by the distance whilst the spectroscopically determined radial velocities are independent of distance. If for example the true distance is underestimated (or overestimated), then so will be the tangential velocities. When inferring the velocity ellipsoid using spectroscopic radial velocities, this will tend to lead to heliocentrically aligned velocity ellipsoids, i.e., the velocity ellipsoids will become elongated (or compressed) towards the solar position. From Fig.~\ref{fig:distances}, we see that using $s=1/\varpi$ overestimates distances therefore will enhance the tangential velocities and cause the velocity ellipsoids to circularise around the Solar position. Notably, the result would be a flattening of the tilt of the velocity ellipsoids at the Solar radius as observed by \citetalias{Hagen2019}.

We use the Bayesian distance estimates derived by \citetalias{Schonrich2019} for our RVS sample. The data set includes corrected parallaxes and parallax uncertainties, which were also revised upwards by \citetalias{Schonrich2019}, and which we use to make quality cuts when applying to this data. Following common practice for parallax-based distance sets, we use $\varpi/\sigma_{\varpi} > 5$  \footnote{\citetalias{Schonrich2019} helpfully provide a $\varpi/\sigma_{\varpi}$ parameter with revised $\sigma_{\varpi}$ which we use to cut on parallax signal-to-noise when applying their distance estimates.}. We select only stars within $5$ kpc of the Sun ($\varpi > 200\,\mu\mathrm{as}$ or $s < 5\,\mathrm{kpc}$ for the Bayesian distances). To remove spurious line-of-sight velocity outliers, we apply $\sigma_{v_r} < 20\,\mathrm{km}\,\mathrm{s}^{-1}$ as well as $|v_r| < 500\,\mathrm{km}\,\mathrm{s}^{-1}$ and further follow the recommendations of \citet{Boubert2019}, which remove stars with less than 4 RVS transits and bright neighbours that can contaminate the measurements.

A concern with \citetalias{Schonrich2019} distance estimates is that the kinematic model prior used to calculate the distances assumed a radially aligned velocity ellipsoid. If this assumption was dominant in the distance inference, our results would be heavily biased towards finding a spherically aligned velocity ellipsoid. We address this concern in two ways. First, we compare \citetalias{Schonrich2019} distance estimates with those found by \citet[][henceforth \citetalias{Anders2019}]{Anders2019} from photo-astrometric distances using the StarHorse pipeline \citep{Queiroz2018}. The potential biases between the \citetalias{Schonrich2019} and \citetalias{Anders2019} distances are very different. The latter set profits in precision from stellar model priors, while it may also inherit biases from the stellar models and have less well-defined distance uncertainties. These two datasets provide an excellent mutual control for remaining biases on either side. To correct for the parallax offset, \citetalias{Anders2019} linearly interpolate as a function of G-band magnitude between the \citet{Lindegren2018} value of $29\mu$as at $G=16.5$ and the $50\mu$as offset found by \citet{Zinn2018} at $G=14$. We place the same cuts to the dataset using \citetalias{Anders2019} distances as described earlier, but using a signal-to-noise cut of $s/\sigma_s>5$ on heliocentric distance rather than parallax. The \citetalias{Anders2019} distance estimates are also in figure ~\ref{fig:distances}. The estimates are similar to \citetalias{Schonrich2019} within 3 kpc where the inference in both methods is dominated by parallax information with low uncertainties. Outside 3 kpc, the distance estimates of \citetalias{Anders2019} are systematically larger by $\sim0.1$ kpc. It is unclear where this disagreement originates from however we find it to be a small enough shift that our results are not significantly affected. In Section \ref{sec:results}, we calculate the tilt for RVS data from StarHorse distances and find it to be consistent with that measured with \citetalias{Schonrich2019}.

\begin{figure}
   \includegraphics[angle=270, width=0.45\textwidth]{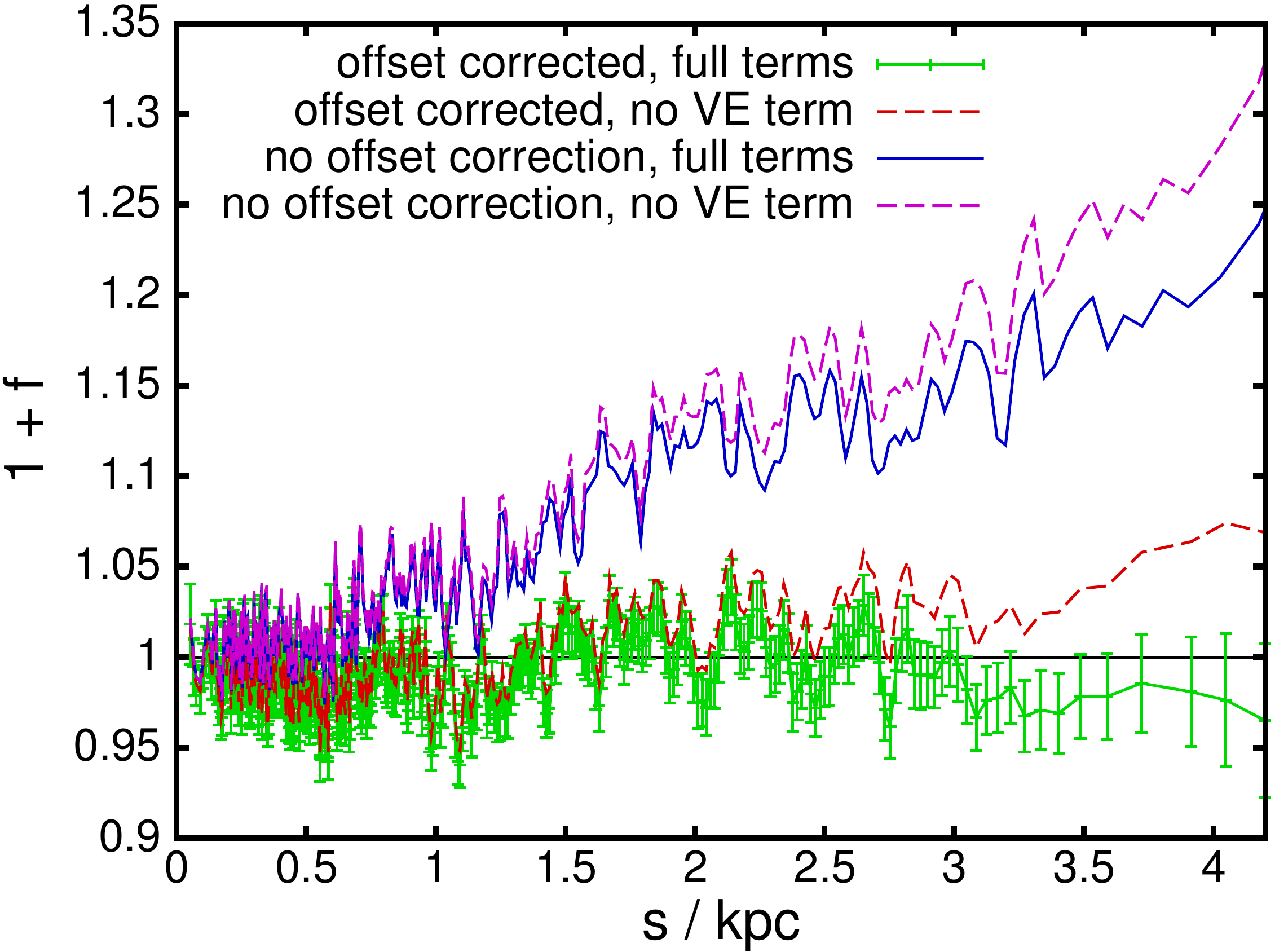}
   \caption{A scan of the Gaia RVS for the fractional distance error $1 + f$ versus distance $s$ with the quality cuts described in \citetalias{Schonrich2019}. Just as in \citetalias{Schonrich2019}, we move a mask of $12000$ stars in steps of $4000$ stars over the sample. The green error bars show the distance statistics after the distance correction, while the solid line shows the statistics when no parallax offset correction is applied. For both values of $\delta_\varpi$, we show with dashed lines the same statistics when we completely remove the velocity ellipsoid correction term, which is equivalent to the wrong assumption that the velocity ellipsoid is cylindrically aligned. The resulting difference overestimates the actual uncertainty, but is still comparably small.}
\label{fig:bias}
\end{figure}

Secondly, to truly quench any remaining uncertainty and to reinforce the use of $54\mu$as offset, we test the effect of the velocity ellipsoid correction terms on distance bias found in \citetalias{Schonrich2019}. This is shown in Fig.~\ref{fig:bias}, where we plot the measured average distance bias versus distance for the \citetalias{Schonrich2019} distances calculated with and without the parallax offset. The dashed lines show the ``measured" distance bias, when we completely remove the velocity ellipsoid correction (which is equivalent to the wrong assumption that the velocity ellipsoid has a perfect cylindrical alignment).

Two things are obvious: i) Even with such a drastic error in assumptions, the change to the distance statistics is less than a third of the overall correction. As a result, the uncertainty in the velocity ellipsoid correction term is more than an order of magnitude smaller than the measured value of the parallax offset in \citetalias{Schonrich2019}. This is also reflected in the systematic uncertainty budget provided by \citetalias{Schonrich2019}. ii) When neglecting the velocity ellipsoid correction term, we actually require a {\bf larger} correction for the parallax offset. As subsequent analysis will show, larger parallax offsets tend to flatten ellipsoids towards the Galactic center and increase the tilt of ellipsoids around and outside the Solar radius. Hence this only strengthens our conclusion that the flattening of the tilt at the solar radius reported by \citetalias{Hagen2019} is driven by biased distance estimates.

\subsection{The LAMOST DR4 and Gaia DR2 Cross-match}

We separately analyse the velocity ellipsoids generated from the combination of 5d phase space information from Gaia DR2 \citep{Brown2018}, together with radial velocities from the LAMOST DR4 value added catalogue \citep{LAMOST2012, Xiang2017}. This enables us to analyse the velocity ellipsoids with an independent catalogue of stars. LAMOST also provides metallicity estimates, which we use to produce halo and thin disc samples by cutting on $\mathrm{[Fe/H]} < -1.5$ and $\mathrm{[Fe/H]}>-0.4$ respectively, as done in \citetalias{Hagen2019}.

We apply the same cuts to this dataset as for RVS, namely  $\varpi/\sigma_{\varpi} > 5$, $\varpi > 200\,\mu\mathrm{as}$, $\sigma_{v_r} < 20\,\mathrm{km}\,\mathrm{s}^{-1}$ and $v_r < 500\,\mathrm{km}\,\mathrm{s}^{-1}$. In the region of overlap between Gaia RVS and LAMOST, we use the radial velocity estimate with the least uncertainty.

We should be cautious of the radial velocities in LAMOST due to the statistical analysis performed by \cite{Schonrich2017}. They determined that the LAMOST radial velocities were offset high by $\sim5\mathrm{km}\,\mathrm{s}^{-1}$. Assuming this offset is global throughout the dataset, it would shift our mean velocities without significantly impacting the velocity dispersions. Hence, we do not include this offset in our analysis.

\section{Method}

To transform from heliocentric to Galactocentric coordinates, we need to fix some Galactic constants. We assume a Solar position~\footnote{The effect of changing the Solar position is investigated in Section 5.1} in cylindrical polar coordinates of $(R_\odot,z_\odot) = (8.27, 0.014)$ kpc \citep[e.g.,][]{Binney1997}. The circular velocity of the Local Standard of Rest is taken as $v_c(R_\odot) = 238 \,\mathrm{km}\,\mathrm{s}^{-1}$ \citep{Schonrich2012}, whilst the Solar peculiar motion is $(U_\odot, V_\odot, W_\odot) = (11.1, 12.24, 7.25)\,\mathrm{km}\,\mathrm{s}^{-1}$ \citep{Schonrich2010}.

We determine the velocity ellipsoid parameters using maximum likelihood estimation on the bivariate Gaussian likelihood function convolved with Gaussian measurement uncertainties similar to previous works~\citep[e.g.,][\citetalias{Hagen2019}]{Bond2010,Evans2016}. We resolve the velocities into Galactocentric spherical polar coordinates ($v_r, v_\theta,v_\phi$) and use a likelihood function 
\begin{equation}
    \log \mathcal{L} = -\frac{1}{2}\log{|{2\pi\Lambda}|} - 
    \frac{1}{2}\sum_i (\boldsymbol{x}_i - \boldsymbol{\mu})^{\mathrm{T}} \Lambda^{-1} (\boldsymbol{x}_i - \boldsymbol{\mu}).
\end{equation}
Here, $\boldsymbol{x}_i = (v_{r.i}, v_{\theta,i})$ are the velocity components of the $i$th star, and $\Lambda = \Sigma + \boldsymbol{C}$, where $\Sigma$ is the velocity covariance matrix in $(v_r,v_\theta)$ and $\boldsymbol{C}$ the measurement uncertainty covariance matrix of the data. The data are binned in a $20 \times 20$ grid of Galactocentric cylindrical polar coordinates $(R, z)$, such that each bin is approximately $500\times500$ pc. For every bin, we analytically calculate the means and covariances of the contained populations without measurement uncertainties. These parameters are used to initialise our likelihood optimization in order to calculate a best fit model with the uncertainties. The algorithm proceeds by optimizing the means and covariances for each bin independently.

For the measurement errors in the RVS sample, we take the standard deviation and correlation parameters for parallaxes, radial velocities and proper motions from the Gaia DR2 dataset. The challenge here is that our likelihood function is inherently Gaussian, whilst, assuming parallax uncertainties are Gaussian, the distance uncertainty distribution is inherently non-Gaussian. When using $1/\varpi$ as our distance estimator, the parallax uncertainty is propagated so we do not assume Gaussian distance uncertainties. However, we do assume Gaussian velocity uncertainties when calculating the likelihood function. When using distance estimates from \citetalias{Schonrich2019}, it is important to use the correct uncertainty distribution. For the purposes of this work, we assume Gaussian distance uncertainties using the second moment of distance given by \citetalias{Schonrich2019} as the variance. For future work, it will be important to understand the impact of the third and fourth moments of distance on our velocity ellipsoids. We also assume here that the distance is uncorrelated with the remaining astrometric parameters. For the LAMOST cross-matched with Gaia sample, we assume that radial velocities are uncorrelated with all the Gaia astrometric parameters.

We determine the parameter posteriors by using the MCMC python package \texttt{emcee} \citep{Foreman-Mackey2013}. We find that initialising walkers in a small ball around our analytically determined parameters allows the chains to converge within 50 iterations. We run 20 walkers for 300 iterations and use the last 150 to calculate our posteriors.

\begin{figure}
   \includegraphics[width=0.5\textwidth]{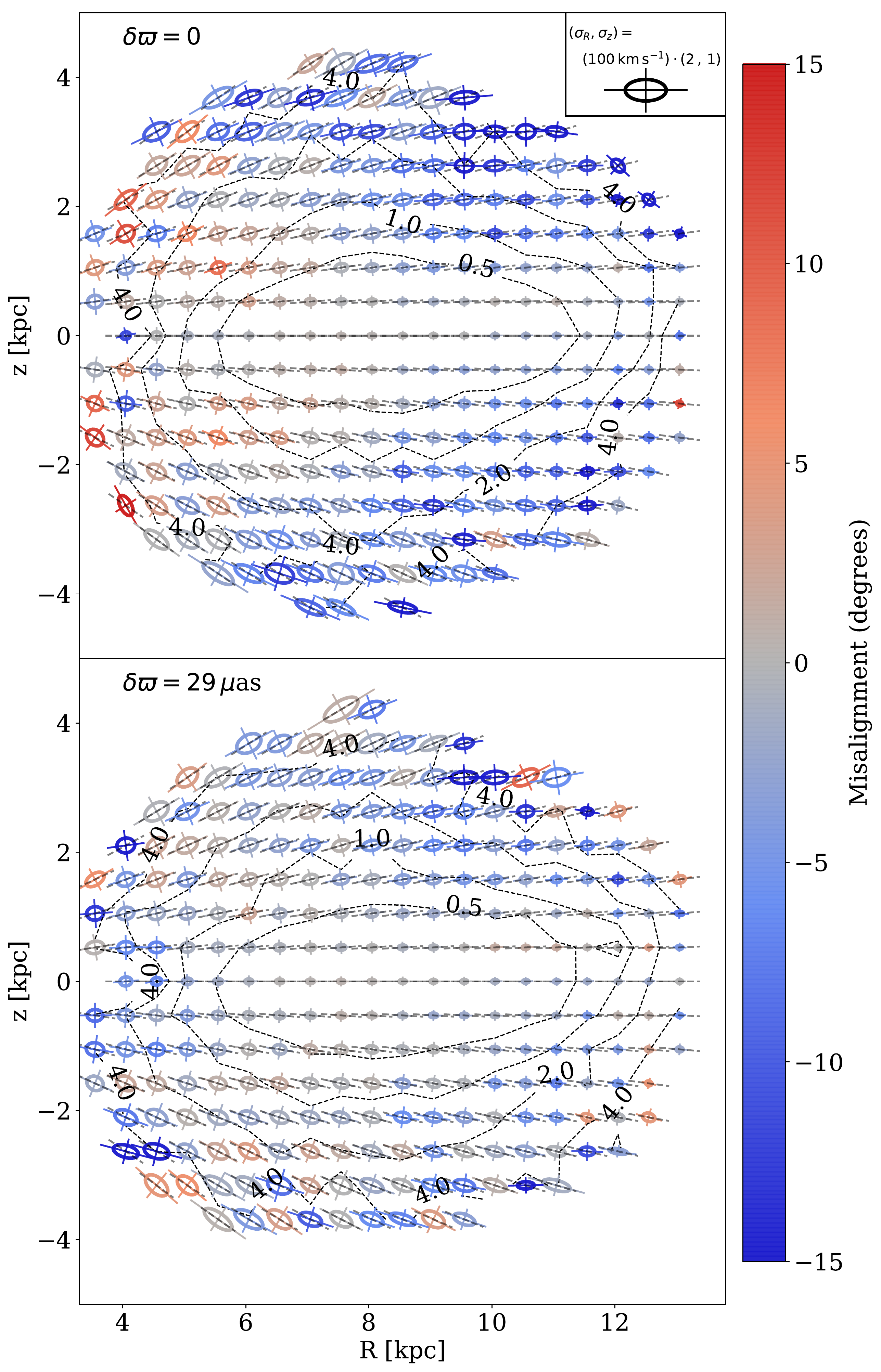}
   \caption{Velocity ellipsoids generated from the Gaia RVS DR2 dataset with different treatments of parallax bias. The size of the ellipsoid is proportional to the value of the velocity dispersion in each bin. The short-dashed lines correspond to the orientation of a spherically aligned velocity dispersion, while the colour bar gives the deviation in degrees of the velocity ellipsoid orientation from this spherical alignment, with blue indicating a flattening and red an over-tilting towards the disk. The black dashed lines show contours of misalignment uncertainty. Top: Using distance as $1/\varpi$ with no parallax offset correction. Bottom: Distance as $1/\varpi$ with $29\,\mu$as parallax correction.}
              \label{fig:rvs_ellipsoids_inversep}
\end{figure}

\begin{figure*}
\centering
   \includegraphics[width=\textwidth]{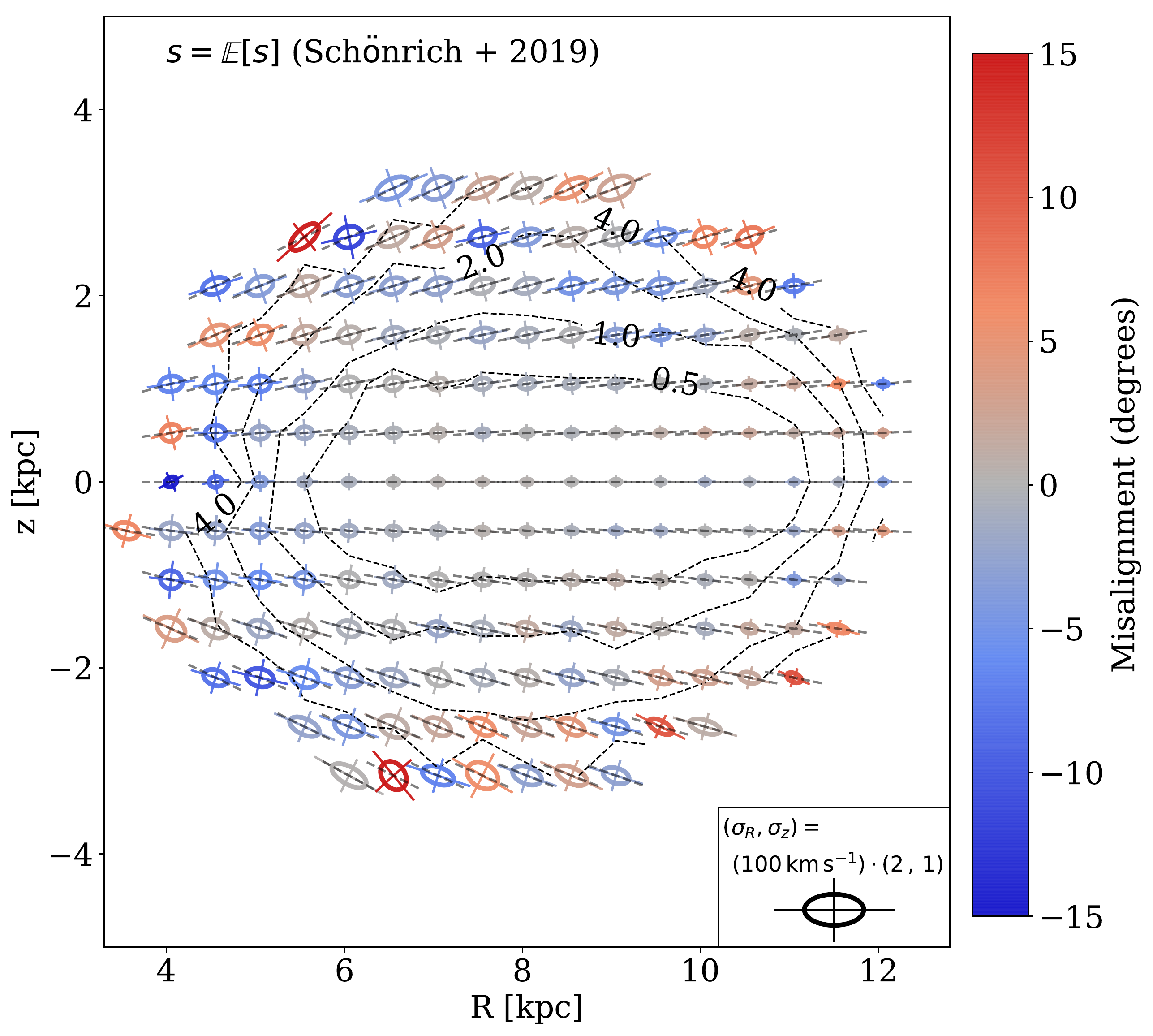}
   \caption{Velocity ellipsoids generated from the Gaia RVS DR2 with Bayesian distance estimates from \citetalias{Schonrich2019} which include a parallax offset correction of $54\,\mu$as. This figure can be compared to Fig.~\ref{fig:rvs_ellipsoids_inversep} which make inferior assumptions as to the distance estimates. Black dashed contours give the ellipsoid orientation uncertainty for 0.5$^\circ$, 1$^\circ$, 2$^\circ$ and 4$^\circ$ respectively. Note that the artificial transition from spherical to cylindrical alignment at the Solar circle visible in the upper panel of Fig.~\ref{fig:rvs_ellipsoids_inversep} has been removed.}
\label{fig:rvs_ellipsoids_sch}
\end{figure*}

\section{Results}
\label{sec:results}
\subsection{The Gaia DR2 RVS sample}

For our analysis of the Gaia RVS sample, we compute the velocity ellipsoids for three different assumptions to show the effects of distance errors:
\begin{enumerate}
\item without any parallax correction and using $s = 1/\varpi$, 
\item with a parallax correction of $29\,\mu\mathrm{as}$ and using $s = 1/\varpi$, 
\item with the Bayesian distance estimates from \citetalias{Schonrich2019}, which use a parallax correction of
$54\,\mu\mathrm{as}$.
\end{enumerate}
Our total sample sizes after applying cuts are 5\,375\,902, 5\,499\,054, and 5\,221\,912 respectively. The velocity ellipsoids produced using assumptions (i) and (ii) are shown in Fig.~\ref{fig:rvs_ellipsoids_inversep}, whilst those produced using (iii) are given greater prominence in Fig.~\ref{fig:rvs_ellipsoids_sch}. We only show ellipsoids in bins with greater than 30 stars, as these still provide clean results and allow us to view the distribution out to greater distances.

In the top panel of Fig.~\ref{fig:rvs_ellipsoids_inversep}, we recover Figure 2 of \citetalias{Hagen2019}. We see the same transition from approximate spherical to cylindrical alignment across the Solar radius. We note that our results are somewhat more noisy, since we have not augmented our data-set with spectroscopic catalogues and so our sample is about 75\% of the size of \citetalias{Hagen2019}. This effect is consistent with overestimates of the distances, and hence tangential velocities, as already discussed in Section 2.1. The bottom panel of Fig.~\ref{fig:rvs_ellipsoids_inversep} shows the same results with a $29\,\mu\mathrm{as}$ correction. The behaviour of the velocity ellipsoid is now much more consistent throughout the meridional plane, without the awkward transition from spherical to cylindrical alignment at the Solar circle. However, of course this correction is conservative and not physically motivated for stars in the RVS sample.

Fig.~\ref{fig:rvs_ellipsoids_sch} uses the Bayesian distance estimates from \citetalias{Schonrich2019} and is the centrepiece of our results. We note that the ellipsoids do not extend out as far as in the previous plots. The reason for this is that \citetalias{Schonrich2019} also revise the parallax uncertainty upwards. As a consequence, when cutting on parallax uncertainty $\varpi/\sigma_\varpi>5$, we remove more stars, particularly at large distances. Those bins which are no longer included do not contain a requisite number of stars for us to plot the ellipsoids. We do observe a slight deviation of the spherical alignment of the velocity ellipsoids at low elevation towards inner radii, tending to cylindrical alignment. This is likely the effect of the contribution of the baryonic disc to the gravitational potential. The same effect can be seen in the velocity ellipsoids of RR Lyrae in the halo in \cite{Wegg2018}, although most of the effect in their analysis occurs within $4$ kpc of the Galactic centre, outside of which the velocity ellipsoids appear to be spherically aligned.

Notice that the size of the velocity ellipsoids increases with elevation above and below the plane. This is caused by the inclusion of three populations of stars, belonging to the thin disc, thick disc and halo. It is interesting to look at the populations separately, and for this we turn to the LAMOST and Gaia cross-matched sample, which has spectroscopic metallicities.

\begin{figure*}
\centering
   \includegraphics[width=\textwidth]{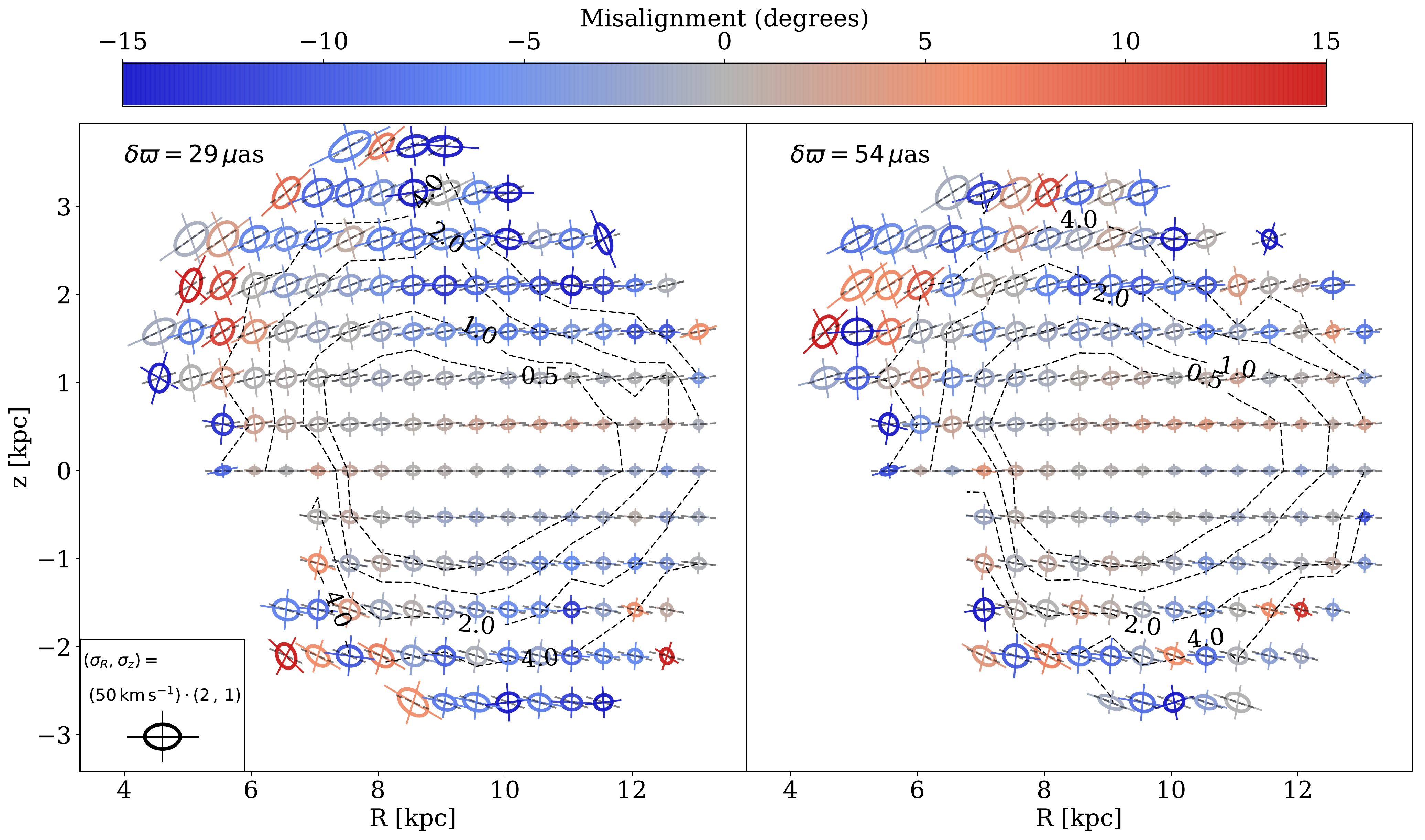}
   \caption{Velocity ellipsoids generated from Gaia DR2 cross-matched with LAMOST with [Fe/H] $>-0.4$ producing a thin disc sample. As usual, the size of the ellipse is related to the value of the velocity dispersion in the given spatial bin. The short dashed lines correspond to the direction of spherical alignment. The colour corresponds to the deviation in degrees of the velocity ellipsoid orientation from spherical alignment. In other words, grey implies spherical alignment whilst blue implies tending towards cylindrical alignment. The black dashed contour shows the misalignment uncertainty. We use $1/\varpi$ as a distance estimator but with $29\,\mu\mathrm{as}$ correction (left panel) and $54\,\mu\mathrm{as}$ correction (right panel). These bracket the range of possibilities, as the former overestimates and the later underestimates the true distances.}
\label{fig:lamost_highfeh_ellipsoids}%
\end{figure*}
\begin{figure*}
\centering
   \includegraphics[width=0.9\textwidth]{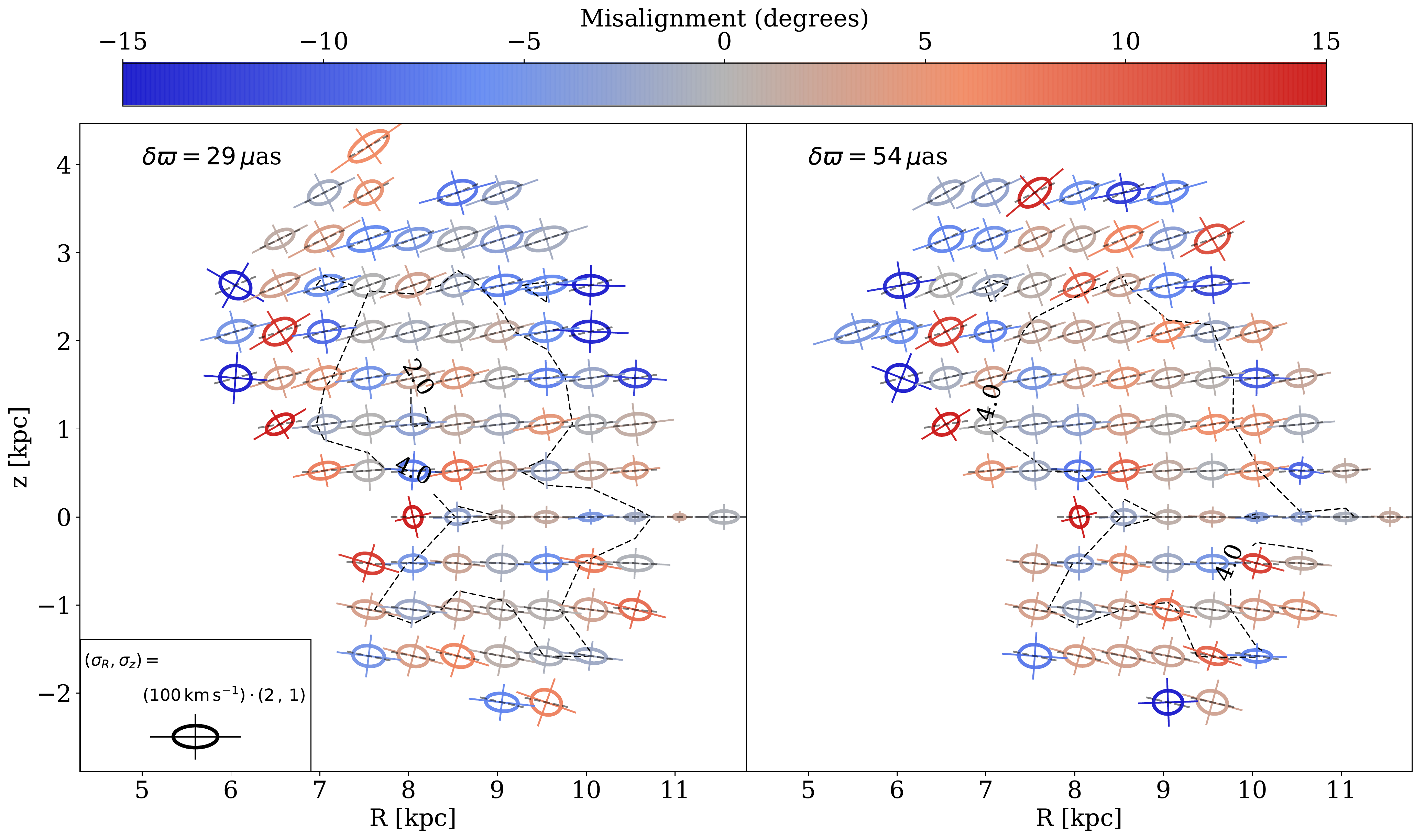}
   \caption{As Fig.~\ref{fig:lamost_highfeh_ellipsoids}, but for the halo sample obtained from Gaia DR2 cross-matched with LAMOST with [Fe/H] $<-1.5$.}
              \label{fig:lamost_lowfeh_ellipsoids}%
\end{figure*}

\subsection{The LAMOST DR4 and Gaia DR2 Crossmatch}

Without Bayesian distances for this sample, we use $s = 1/\varpi$ as our estimator with parallax corrections $29\mu\mathrm{as}$ and $54\mu\mathrm{as}$. We expect these to overestimate and underestimate distances respectively, as indicated by Fig.~\ref{fig:distances}. Therefore, our results on the tilt of the velocity ellipsoid merely bracket the range of possibilities.

We split the sample into two separate populations, [Fe/H] $>-0.4$ as a thin disc sample and [Fe/H] $<-1.5$ as a halo sample. Neither sample is completely pure, as the metallicity cuts only approximately separate populations. After applying the cuts, our halo samples contain 18\,424 and 19\,661 stars for $29\mu$as and $54\mu$as corrections respectively and the thin disc samples contain 2\,286\,528 and 2\,306\,729 stars.

In Fig.~\ref{fig:lamost_highfeh_ellipsoids}, we present results for the thin disc sample. In the left plot, the flattening of the tilt is still strong for the $29\,\mu$as correction, with cylindrical alignment particularly prevalent at elevations above 2 kpc from the plane. 
In the right plot, with a $54\,\mu$as correction, the majority of this signal has been removed. However, there appears to be a small but significant deviation from spherical alignment remaining for heights $|z|\sim2.5$ kpc. It is suggestive that there the thin disc population may not be exactly spherically aligned.

The results for the low metallicity halo sample are given in Fig.~\ref{fig:lamost_lowfeh_ellipsoids}. This contains a much smaller number of stars, which allows us fewer bins and causes the results to appear more noisy. However, in the left plot, with the conservative $29\mu$as correction, almost cylindrical alignment can be seen for $R\sim10$ kpc and $z\sim2$ kpc which is completely removed in the right hand plot for the $54\,\mu$as over-correction. We also note here that the scales of the velocity dispersions are much more consistent across elevations which demonstrates the effect of selection of the halo sample with only small impurities.


\begin{figure}
   \includegraphics[width=0.45\textwidth]{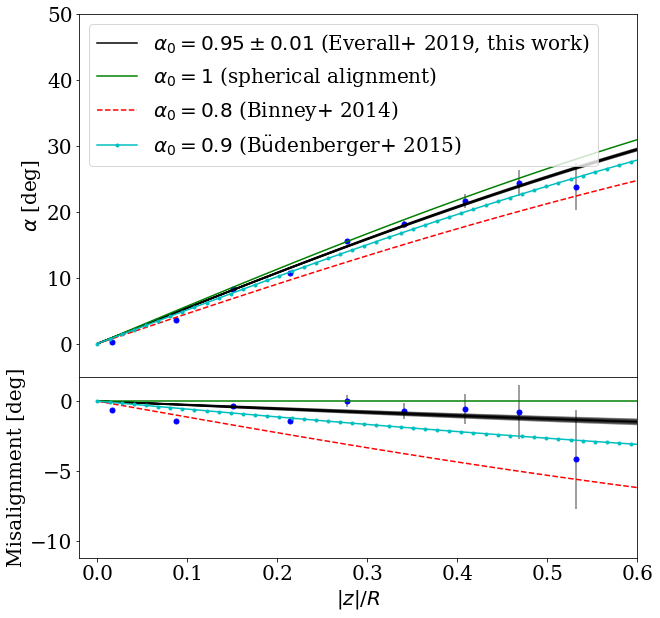}
   \caption{The upper panel shows fits for the tilt of the velocity ellipsoid using eq.~(\ref{eq:alpha}). The blue points provide the posterior means and uncertainties of ellipsoid inclinations in $|z|/R$ bins. Perfect spherical alignment corresponds to the green line, whereas the black line is our result from the Gaia RVS sample with distances from \citet{Schonrich2019}. For comparison, we also show recent fits from \citet{Binney2014} (red) and \citet{Budenbender2015} (pale blue). Notice that the binned datapoints show a transition from below to above the best fit line, as the disc potential becomes less dominant. The lower panel shows the deviation from spherical alignment.}
              \label{fig:tilt}
\end{figure}

\section{The Tilt of the Velocity Ellipsoid}

\citet{Binney2014} and \citet{Budenbender2015} introduced and exploited a compact way to summarize results on the tilt of the velocity ellipsoids. They used a model in which the angle between the Galactic plane and the direction of the longest axis of the velocity ellipsoid is
\begin{equation}
\alpha = \alpha_0\,\arctan{|z| / R}.
\label{eq:alpha}
\end{equation}
They fitted the binned data to the model to determine the best fit $\alpha_0$ parameter. A result of $\alpha_0 = 1$ implies exact spherical alignment, whilst $\alpha_0 < 1$ means that the ellipsoids are tilted towards cylindrical alignment. 

We perform a least squares regression on all bins with $n_{\mathrm{stars}}>5$ as these still contain valuable information about ellipsoid alignment although with large uncertainties\footnote{In Section \ref{sec:results}, we only use bins with $n_{\mathrm{stars}}>30$ because the scatter in less populated bins make the ellipsoid plots appear untidy and muddied the trends in behaviour.}. Bins with fewer stars are almost randomly aligned. For the Gaia RVS sample with \citetalias{Schonrich2019} distances, we acquire a tilt value of $\alpha_0 = 0.952\pm0.007$. This is in significant disagreement with $\alpha_0 \sim 0.8$ determined in \citet{Binney2014} from the local RAVE stars~\citep{Steinmetz2006}. It is in reasonable agreement with \citet{Budenbender2015}, who found a value of $0.90 \pm 0.04$ using the Segue G dwarf sample. As discussed in Section~\ref{sec:data}, we also calculate this parameter for the distance estimates of \citetalias{Anders2019} with the RVS sample and retrieve $\alpha_0=0.956\pm0.006$, in remarkably good agreement with the estimate from \citetalias{Schonrich2019} distances.

We see no physical reason why this parameter should be constant across all populations of stars and in all parts of the Galaxy. Under the hypothesis that tilt of the velocity ellipsoids is controlled at least in part by the contribution of the baryonic disc to the potential, we anticipate that $\alpha_0$ should be lowest near the plane and tend towards $1$ at high elevation. We also suggest that the flattening of the tilt should be more extreme in the inner radii. To test this hypothesis, we compute $\alpha_0$ for subsets of our velocity ellipsoids. We find that for $|z|<2\,\mathrm{kpc}$, $\alpha_0 = 0.950 \pm 0.007$ whilst for $|z|>2\,\mathrm{kpc}$, $\alpha_0 = 0.966 \pm 0.018$. We also find that at $R<7\,\mathrm{kpc}$, $\alpha_0 = 0.917 \pm 0.013$ whilst for $R>7\,\mathrm{kpc}$, $\alpha_0 = 0.963 \pm 0.007$. This is consistent with the hypothesis that the effects of the disc potential are driving much of the deviation from spherical alignment.

We also look at the tilt at large radii and high elevation. For $|z|>2\,\mathrm{kpc}$ and $R>7\,\mathrm{kpc}$, we retrieve the result $\alpha_0=0.986\pm0.020$, which is consistent with spherical alignment. This is in good agreement with a number of studies of the velocity ellipsoids of halo stars in SDSS \citep{Smith2009,Bond2010,Evans2016}, as well as the recent work of \citet{Wegg2018} who determined that the kinematics of the RR Lyrae in the halo, extracted from Gaia DR2, imply a spherically symmetric halo potential.

In Fig.~\ref{fig:tilt}, we show the fit of the tilt of the velocity ellipsoids as a function of $|z|/R$. The green solid line shows the expected trend for spherical alignment ($\alpha_0=1$). We plot our best fit, as well as the earlier results from \cite{Binney2014} and \cite{Budenbender2015}. The blue points are the posterior means of uncertainties of ellipsoid inclinations and misalignments in $|z|/R$ bins. Notice that the binned datapoints show an interesting pattern with respect to the best fit. The datapoints with high $|z|$ mostly lie just above the best fit, those with low $|z|$ lie just below. This trend suggests that the deviations from spherical alignment are induced by the disc potential.

We also compare ellipsoids above and below the plane. We find that above the disc $\alpha_0=0.964\pm0.009$, whilst below the plane, $\alpha_0=0.940\pm0.009$, showing $2\sigma$ disagreement. However, this asymmetry is far more stark when separating in-plane from high elevation contributions. Considering only ellipsoids within 1kpc of the plane, we find that $\alpha_0=0.989\pm0.014$ above and $0.888\pm0.013$ below which has a $5\sigma$ difference. Conversely outside 1kpc, $\alpha_0=0.94\pm0.01$ and $0.99\pm0.01$ above and below respectively, in $3\sigma$ disagreement and opposite to the in-plane difference. 

For an axisymmetric equilibrium that is reflexion symmetric about the Galactic plane, results above and below the plane should be consistent. This apparent discrepancy particularly in the disk may be caused by substructure and streams, buckling of the Galactic bar~\citep{Sa13}, or by the effects of bending modes in the disc~\citep[e.g.][]{Gomez2013,Wi3,Xu2015,La19}, or by unrecognized systematics in the data.

We analyse the thin disc and halo samples generated from the Gaia-LAMOST cross-match. For the disc sample, we recover $\alpha=0.909\pm0.008$ for the $29\mu$as correction, which becomes $\alpha=1.038\pm0.008$ for the $54\,\mu$as correction. As anticipated, this straddles the RVS results demonstrating the effect of overestimating and underestimating the distances. The same effect is present in the halo sample with $\alpha=0.927\pm0.035$ and $\alpha=1.063\pm0.036$ for corrections of $29\,\mu$as and $54\,\mu$as respectively. 

\subsection{The Solar position}

In the analysis, we assumed a Solar distance to the Galactic center of $R_{\odot}=8.27$ \citep{Binney1997} and neglected uncertainties on this estimate. This is mainly to ease comparison with earlier work, especially \citetalias{Hagen2019}. Recently, the \citet{GRAVITY2018} reported a high precision distance to Sagittarius A* of $8.127\pm0.031$ kpc, which is smaller than our assumed value.

Adjusting the Solar position with respect to the Galactic centre does not change the properties of velocity ellipsoids in Cartesian coordinates. The only impact is that we now calculate the misalignment with respect to a new central point in the Galaxy.

For this change in $R_\odot$, the shift in misalignment is small. In the most extreme cases of velocity ellipsoids at ($|z|\sim2$, $R\sim4$) kpc, the misalignment is reduced by $0.84^\circ$ which falls well within our uncertainties. On average, across all our ellipsoid positions, the induced flattening is $0.33^\circ$. The effect on any individual ellipsoid is negligible.

However, a change in $R_\odot$ induces a coherent shift in all ellipsoid misalignments, and so there is a somewhat larger effect on our inference of the tilt normalization parameter, $\alpha_0$. We find that using $R_\odot = 8.127$ kpc, the full RVS sample generates a tilt parameter of $\alpha_0=0.953 \pm 0.007$. This shift is still within the original uncertainties. Similar calculations for sub-samples of the ellipsoids prove even less significant due to their increased uncertainties.

\section{Conclusions}
\label{sec:conclude}

The tilt of the velocity ellipsoid of local stars is important for several reasons. First, determinations of the local dark matter density are usually based on the vertical kinematics of stars. The gravitational potential is inferred from the Jeans equations or distribution functions, a calculation known to be sensitive to the tilt of the velocity ellipsoid~\citep[e.g.][]{Si16, Sivertsson2018}. Secondly, the heating processes that thicken discs include scattering by in-plane spiral arms and by giant molecular clouds. These scattering processes can produce different signatures in the tilt of the thin disc velocity ellipsoid~\citep[e.g.,][]{Se14}. Thirdly, the alignment can give direct information on the potential in some instances~\citep[e.g.,][]{Edd15,Bi11}. For example, the halo stars are believed to be close to spherical alignment, as judged by a number of earlier studies of SDSS star samples~\citep[e.g.,][]{Bond2010}. Exact spherical alignment implies a spherically symmetric force field~\citep{Smith2009,An16}.

The Gaia Radial Velocity Spectrometer (RVS) sample 
comprises 7\,224\,631 stars with full phase space coordinates. The main hurdle to overcome in exploiting this dataset to study the tilt is the accurate and unbiased conversion of parallaxes $\varpi$ to heliocentric distances $s$. We find that the Bayesian distances of \citet{Schonrich2019}, which incorporate a parallax offset of $54\,\mu\mathrm{as}$, give reliable results. We have checked that substitution of photo-astrometric distances from \citet{Anders2019} using the StarHorse pipeline
gives consistent results. However, use of the reciprocal of parallax as a distance estimator leads to artefacts in the behaviour of the inferred velocity ellipsoids and this practise should be deprecated.

The Gaia RVS sample is consistent with nearly spherical alignment. The tilt is accurately described by the relation $\alpha = (0.952\pm 0.007)\arctan (|z|/R)$. If the normalising constant were unity, then this would imply exact alignment with spherical polars. Our result is pleasingly close to that found by \citet{Budenbender2015} from the Segue G dwarf stars in the Solar neighbourhood. If the sample is restricted to stars at large Galactocentric radii, or great distances above or below the plane, then the alignment becomes still closer to spherical. The data support the conjecture that any deviation from spherical alignment of the velocity ellipsoids is caused by the gravitational potential of the disc. Such deviations occur at low $|z|$ and close to the Galactic center, whilst at $|z|>2$ kpc and $R>7$ kpc the ellipsoids are consistent with spherical alignment. 

Subsamples from Gaia DR2 cross-matched with LAMOST enable us to study the disc and halo populations separately. Even though Bayesian distances are not available for all these stars, we can bracket the tilt of the velocity ellipsoids by making assumptions that underestimate and overestimate the heliocentric distances. For thin disc stars, we find $\alpha = (0.909-1.038)\arctan (|z|/R)$ and for halo stars $\alpha = (0.927-1.063)\arctan (|z|/R)$. Both populations are close to spherical alignment, with the only real deviations occurring in the inner Galaxy near the Galactic plane.

Here, we have studied only the orientation of the velocity ellipsoids as seen by Gaia. Our results have important implications for the local dark matter density, for which treatment of the tilt term is a major source of the uncertainty. We plan to attack this problem in a forthcoming publication.

\section*{Acknowledgments}
AE thanks the Science and Technology Facilities Council of the United Kingdom for financial support. This work was partly performed at the Oxford Gaia sprint, and we thank Douglas Boubert and Payel Das for organising the meeting. The work was completed at the Kavli Institute for Theoretical Physics, Santa Barbara. NWE and VB acknowledge support in part by the National Science Foundation under Grant No. NSF PHY-1748958.



\bibliographystyle{mnras}
\bibliography{./LocalVelocities}

\begin{thebibliography}{}
\makeatletter
\relax
\def\mn@urlcharsother{\let\do\@makeother \do\$\do\&\do\#\do\^\do\_\do\%\do\~}
\def\mn@doi{\begingroup\mn@urlcharsother \@ifnextchar [ {\mn@doi@}
  {\mn@doi@[]}}
\def\mn@doi@[#1]#2{\def\@tempa{#1}\ifx\@tempa\@empty \href
  {http://dx.doi.org/#2} {doi:#2}\else \href {http://dx.doi.org/#2} {#1}\fi
  \endgroup}
\def\mn@eprint#1#2{\mn@eprint@#1:#2::\@nil}
\def\mn@eprint@arXiv#1{\href {http://arxiv.org/abs/#1} {{\tt arXiv:#1}}}
\def\mn@eprint@dblp#1{\href {http://dblp.uni-trier.de/rec/bibtex/#1.xml}
  {dblp:#1}}
\def\mn@eprint@#1:#2:#3:#4\@nil{\def\@tempa {#1}\def\@tempb {#2}\def\@tempc
  {#3}\ifx \@tempc \@empty \let \@tempc \@tempb \let \@tempb \@tempa \fi \ifx
  \@tempb \@empty \def\@tempb {arXiv}\fi \@ifundefined
  {mn@eprint@\@tempb}{\@tempb:\@tempc}{\expandafter \expandafter \csname
  mn@eprint@\@tempb\endcsname \expandafter{\@tempc}}}

\bibitem[\protect\citeauthoryear{Abolfathi et~al.,}{Abolfathi
  et~al.}{2018}]{Abolfathi2018}
Abolfathi B.,  et~al., 2018, \mn@doi [\apjs] {10.3847/1538-4365/aa9e8a}, 235,
  42

\bibitem[\protect\citeauthoryear{Adelmam-McCarthy et~al.,}{Adelmam-McCarthy
  et~al.}{2008}]{AdelmanMcCarthy2008}
Adelmam-McCarthy J.~K.,  et~al., 2008, \mn@doi [\apjs] {10.1086/524984}, 175,
  297

\bibitem[\protect\citeauthoryear{{An} \& {Evans}}{{An} \& {Evans}}{2016}]{An16}
{An} J.,  {Evans} N.~W.,  2016, \mn@doi [\apj] {10.3847/0004-637X/816/1/35},
  \href {https://ui.adsabs.harvard.edu/abs/2016ApJ...816...35A} {816, 35}

\bibitem[\protect\citeauthoryear{{Anders} et~al.,}{{Anders}
  et~al.}{2019}]{Anders2019}
{Anders} F.,  et~al., 2019, arXiv e-prints, \href
  {https://ui.adsabs.harvard.edu/abs/2019arXiv190411302A} {p. arXiv:1904.11302}

\bibitem[\protect\citeauthoryear{Bailer-Jones}{Bailer-Jones}{2015}]{Bailer-Jones2015}
Bailer-Jones C. A.~L.,  2015, \mn@doi [\pasp] {10.1086/683116}, 127, 994

\bibitem[\protect\citeauthoryear{{Bailer-Jones}, {Rybizki}, {Fouesneau},
  {Mantelet}  \& {Andrae}}{{Bailer-Jones} et~al.}{2018}]{Ba18}
{Bailer-Jones} C.~A.~L.,  {Rybizki} J.,  {Fouesneau} M.,  {Mantelet} G.,
  {Andrae} R.,  2018, \mn@doi [\aj] {10.3847/1538-3881/aacb21}, \href
  {https://ui.adsabs.harvard.edu/abs/2018AJ....156...58B} {156, 58}

\bibitem[\protect\citeauthoryear{{Binney} \& {McMillan}}{{Binney} \&
  {McMillan}}{2011}]{Bi11}
{Binney} J.,  {McMillan} P.,  2011, \mn@doi [\mnras]
  {10.1111/j.1365-2966.2011.18268.x}, \href
  {http://adsabs.harvard.edu/abs/2011MNRAS.413.1889B} {413, 1889}

\bibitem[\protect\citeauthoryear{Binney, Gerhard  \& Spergel}{Binney
  et~al.}{1997}]{Binney1997}
Binney J.,  Gerhard O.,   Spergel D.,  1997, \mn@doi [\mnras]
  {10.1093/mnras/288.2.365}, 288, 365

\bibitem[\protect\citeauthoryear{Binney et~al.,}{Binney
  et~al.}{2014}]{Binney2014}
Binney J.,  et~al., 2014, \mn@doi [\mnras] {10.1093/mnras/stt2367}, 439, 1231

\bibitem[\protect\citeauthoryear{Bond et~al.,}{Bond et~al.}{2010}]{Bond2010}
Bond N.~A.,  et~al., 2010, \mn@doi [\apj] {10.1088/0004-637X/716/1/1}, 716, 1

\bibitem[\protect\citeauthoryear{{Boubert} et~al.,}{{Boubert}
  et~al.}{2019}]{Boubert2019}
{Boubert} D.,  et~al., 2019, \mn@doi [\mnras] {10.1093/mnras/stz253}, \href
  {https://ui.adsabs.harvard.edu/abs/2019MNRAS.486.2618B} {486, 2618}

\bibitem[\protect\citeauthoryear{Brown et~al.,}{Brown et~al.}{2018}]{Brown2018}
Brown A. G.~A.,  et~al., 2018, \mn@doi [\aap] {10.1051/0004-6361/201833051},
  616, A1

\bibitem[\protect\citeauthoryear{B{\"{u}}denbender, van~de Ven  \&
  Watkins}{B{\"{u}}denbender et~al.}{2015}]{Budenbender2015}
B{\"{u}}denbender A.,  van~de Ven G.,   Watkins L.~L.,  2015, \mn@doi [\mnras]
  {10.1093/mnras/stv1314}, 452, 956

\bibitem[\protect\citeauthoryear{Cui et~al.,}{Cui et~al.}{2012}]{LAMOST2012}
Cui X.-Q.,  et~al., 2012, \mn@doi [Research in Astronomy and Astrophysics]
  {10.1088/1674-4527/12/9/003}, 12, 1197

\bibitem[\protect\citeauthoryear{{Eddington}}{{Eddington}}{1915}]{Edd15}
{Eddington} A.~S.,  1915, \mn@doi [\mnras] {10.1093/mnras/76.1.37}, \href
  {http://adsabs.harvard.edu/abs/1915MNRAS..76...37E} {76, 37}

\bibitem[\protect\citeauthoryear{Evans, Sanders, Williams, An, Lynden-Bell  \&
  Dehnen}{Evans et~al.}{2016}]{Evans2016}
Evans N.~W.,  Sanders J.~L.,  Williams A.~A.,  An J.,  Lynden-Bell D.,   Dehnen
  W.,  2016, \mn@doi [\mnras] {10.1093/mnras/stv2729}, 456, 4506

\bibitem[\protect\citeauthoryear{Foreman-Mackey, Hogg, Lang  \&
  Goodman}{Foreman-Mackey et~al.}{2013}]{Foreman-Mackey2013}
Foreman-Mackey D.,  Hogg D.~W.,  Lang D.,   Goodman J.,  2013, \mn@doi [\pasp]
  {10.1086/670067}, 125, 306

\bibitem[\protect\citeauthoryear{{G{\'o}mez}, {Minchev}, {O'Shea}, {Beers},
  {Bullock}  \& {Purcell}}{{G{\'o}mez} et~al.}{2013}]{Gomez2013}
{G{\'o}mez} F.~A.,  {Minchev} I.,  {O'Shea} B.~W.,  {Beers} T.~C.,  {Bullock}
  J.~S.,   {Purcell} C.~W.,  2013, \mn@doi [\mnras] {10.1093/mnras/sts327},
  \href {https://ui.adsabs.harvard.edu/abs/2013MNRAS.429..159G} {429, 159}

\bibitem[\protect\citeauthoryear{{Gravity Collaboration} et~al.,}{{Gravity
  Collaboration} et~al.}{2018}]{GRAVITY2018}
{Gravity Collaboration} et~al., 2018, \mn@doi [\aap]
  {10.1051/0004-6361/201833718}, \href
  {http://adsabs.harvard.edu/abs/2018A%26A...615L..15G} {615, L15}

\bibitem[\protect\citeauthoryear{Hagen, Helmi, de Zeeuw  \& Posti}{Hagen
  et~al.}{2019}]{Hagen2019}
Hagen J. H.~J.,  Helmi A.,  de Zeeuw P.~T.,   Posti L.,  2019, arXiv:1902.05268

\bibitem[\protect\citeauthoryear{{Katz} et~al.,}{{Katz} et~al.}{2019}]{Ka19}
{Katz} D.,  et~al., 2019, \mn@doi [\aap] {10.1051/0004-6361/201833273}, \href
  {http://adsabs.harvard.edu/abs/2019A%26A...622A.205K} {622, A205}

\bibitem[\protect\citeauthoryear{Khan et~al.,}{Khan et~al.}{2019}]{Khan2019}
Khan S.,  et~al., 2019, arXiv:1904.05676

\bibitem[\protect\citeauthoryear{Kunder et~al.,}{Kunder
  et~al.}{2017}]{Kunder2017}
Kunder A.,  et~al., 2017, \mn@doi [\aj] {10.3847/1538-3881/153/2/75}, 153, 75

\bibitem[\protect\citeauthoryear{{Laporte}, {Minchev}, {Johnston}  \&
  {G{\'o}mez}}{{Laporte} et~al.}{2019}]{La19}
{Laporte} C.~F.~P.,  {Minchev} I.,  {Johnston} K.~V.,   {G{\'o}mez} F.~A.,
  2019, \mn@doi [\mnras] {10.1093/mnras/stz583}, \href
  {http://adsabs.harvard.edu/abs/2019MNRAS.485.3134L} {485, 3134}

\bibitem[\protect\citeauthoryear{Lindegren et~al.,}{Lindegren
  et~al.}{2018}]{Lindegren2018}
Lindegren L.,  et~al., 2018, \mn@doi [\aap] {10.1051/0004-6361/201832727}, 616,
  A2

\bibitem[\protect\citeauthoryear{{Luri} et~al.,}{{Luri}
  et~al.}{2018}]{Luri2018}
{Luri} X.,  et~al., 2018, \mn@doi [\aap] {10.1051/0004-6361/201832964}, \href
  {https://ui.adsabs.harvard.edu/abs/2018A&A...616A...9L} {616, A9}

\bibitem[\protect\citeauthoryear{{Lutz} \& {Kelker}}{{Lutz} \&
  {Kelker}}{1973}]{Lutz1973}
{Lutz} T.~E.,  {Kelker} D.~H.,  1973, \mn@doi [\pasp] {10.1086/129506}, \href
  {https://ui.adsabs.harvard.edu/abs/1973PASP...85..573L} {85, 573}

\bibitem[\protect\citeauthoryear{{Queiroz} et~al.,}{{Queiroz}
  et~al.}{2018}]{Queiroz2018}
{Queiroz} A.~B.~A.,  et~al., 2018, \mn@doi [\mnras] {10.1093/mnras/sty330},
  \href {https://ui.adsabs.harvard.edu/abs/2018MNRAS.476.2556Q} {476, 2556}

\bibitem[\protect\citeauthoryear{Riess et~al.,}{Riess et~al.}{2018}]{Riess2018}
Riess A.~G.,  et~al., 2018, \mn@doi [\apj] {10.3847/1538-4357/aaadb7}, 855, 136

\bibitem[\protect\citeauthoryear{{Saha}, {Pfenniger}  \& {Taam}}{{Saha}
  et~al.}{2013}]{Sa13}
{Saha} K.,  {Pfenniger} D.,   {Taam} R.~E.,  2013, \mn@doi [\apj]
  {10.1088/0004-637X/764/2/123}, \href
  {http://adsabs.harvard.edu/abs/2013ApJ...764..123S} {764, 123}

\bibitem[\protect\citeauthoryear{Sch{\"{o}}nrich}{Sch{\"{o}}nrich}{2012}]{Schonrich2012}
Sch{\"{o}}nrich R.,  2012, \mn@doi [\mnras] {10.1111/j.1365-2966.2012.21631.x},
  427, 274

\bibitem[\protect\citeauthoryear{Sch{\"{o}}nrich \& Aumer}{Sch{\"{o}}nrich \&
  Aumer}{2017}]{Schonrich2017}
Sch{\"{o}}nrich R.,  Aumer M.,  2017, \mn@doi [\mnras] {10.1093/mnras/stx2189},
  472, 3979

\bibitem[\protect\citeauthoryear{Sch{\"{o}}nrich, Binney  \&
  Dehnen}{Sch{\"{o}}nrich et~al.}{2010}]{Schonrich2010}
Sch{\"{o}}nrich R.,  Binney J.,   Dehnen W.,  2010, \mn@doi [\mnras]
  {10.1111/j.1365-2966.2010.16253.x}, 403, 1829

\bibitem[\protect\citeauthoryear{Sch{\"{o}}nrich, McMillan  \&
  Eyer}{Sch{\"{o}}nrich et~al.}{2019}]{Schonrich2019}
Sch{\"{o}}nrich R.,  McMillan P.,   Eyer L.,  2019, arXiv:1902.02355

\bibitem[\protect\citeauthoryear{Sellwood}{Sellwood}{2014}]{Se14}
Sellwood J.~A.,  2014, \mn@doi [Rev. Mod. Phys.] {10.1103/RevModPhys.86.1}, 86,
  1

\bibitem[\protect\citeauthoryear{{Silverwood}, {Sivertsson}, {Steger}, {Read}
  \& {Bertone}}{{Silverwood} et~al.}{2016}]{Si16}
{Silverwood} H.,  {Sivertsson} S.,  {Steger} P.,  {Read} J.~I.,   {Bertone} G.,
   2016, \mn@doi [\mnras] {10.1093/mnras/stw917}, \href
  {http://adsabs.harvard.edu/abs/2016MNRAS.459.4191S} {459, 4191}

\bibitem[\protect\citeauthoryear{{Sivertsson}, {Silverwood}, {Read}, {Bertone}
  \& {Steger}}{{Sivertsson} et~al.}{2018}]{Sivertsson2018}
{Sivertsson} S.,  {Silverwood} H.,  {Read} J.~I.,  {Bertone} G.,   {Steger} P.,
   2018, \mn@doi [\mnras] {10.1093/mnras/sty977}, \href
  {https://ui.adsabs.harvard.edu/abs/2018MNRAS.478.1677S} {478, 1677}

\bibitem[\protect\citeauthoryear{Smith, Evans  \& An}{Smith
  et~al.}{2009}]{Smith2009}
Smith M.~C.,  Evans N.~W.,   An J.~H.,  2009, \mn@doi [\apj]
  {10.1088/0004-637X/698/2/1110}, 698, 1110

\bibitem[\protect\citeauthoryear{Steinmetz et~al.,}{Steinmetz
  et~al.}{2006}]{Steinmetz2006}
Steinmetz M.,  et~al., 2006, \mn@doi [\aj] {10.1086/506564}, 132, 1645

\bibitem[\protect\citeauthoryear{{Str{\"o}mberg}}{{Str{\"o}mberg}}{1927}]{Stroemberg27}
{Str{\"o}mberg} G.,  1927, \mn@doi [\apj] {10.1086/143045}, \href
  {https://ui.adsabs.harvard.edu/abs/1927ApJ....65..238S} {65, 238}

\bibitem[\protect\citeauthoryear{Wegg, Gerhard  \& Bieth}{Wegg
  et~al.}{2019}]{Wegg2018}
Wegg C.,  Gerhard O.,   Bieth M.,  2019, \mn@doi [arXiv:1806.09635]
  {10.1093/mnras/stz572}

\bibitem[\protect\citeauthoryear{{Williams} et~al.,}{{Williams}
  et~al.}{2013}]{Wi3}
{Williams} M.~E.~K.,  et~al., 2013, \mn@doi [\mnras] {10.1093/mnras/stt1522},
  \href {http://adsabs.harvard.edu/abs/2013MNRAS.436..101W} {436, 101}

\bibitem[\protect\citeauthoryear{{Xiang} et~al.,}{{Xiang}
  et~al.}{2017}]{Xiang2017}
{Xiang} M.-S.,  et~al., 2017, \mn@doi [\mnras] {10.1093/mnras/stx129}, \href
  {https://ui.adsabs.harvard.edu/abs/2017MNRAS.467.1890X} {467, 1890}

\bibitem[\protect\citeauthoryear{{Xu}, {Newberg}, {Carlin}, {Liu}, {Deng},
  {Li}, {Sch{\"o}nrich}  \& {Yanny}}{{Xu} et~al.}{2015}]{Xu2015}
{Xu} Y.,  {Newberg} H.~J.,  {Carlin} J.~L.,  {Liu} C.,  {Deng} L.,  {Li} J.,
  {Sch{\"o}nrich} R.,   {Yanny} B.,  2015, \mn@doi [\apj]
  {10.1088/0004-637X/801/2/105}, \href
  {https://ui.adsabs.harvard.edu/abs/2015ApJ...801..105X} {801, 105}

\bibitem[\protect\citeauthoryear{Xu, Zhang, Reid, Zheng  \& Wang}{Xu
  et~al.}{2019}]{Xu2019}
Xu S.,  Zhang B.,  Reid M.~J.,  Zheng X.,   Wang G.,  2019, arXiv:1903.04195

\bibitem[\protect\citeauthoryear{Yanny et~al.,}{Yanny et~al.}{2009}]{Yanny2009}
Yanny B.,  et~al., 2009, \mn@doi [\aj] {10.1088/0004-6256/137/5/4377}, 137,
  4377

\bibitem[\protect\citeauthoryear{Zinn, Pinsonneault, Huber  \& Stello}{Zinn
  et~al.}{2018}]{Zinn2018}
Zinn J.~C.,  Pinsonneault M.~H.,  Huber D.,   Stello D.,  2018,
  arXiv:1805.02650

\makeatother
\end{thebibliography}

\label{lastpage}
\end{document}